\documentclass[aps,prb,twocolumn,superscriptaddress,showpacs,floatfix]{revtex4}
\usepackage{graphicx}
\usepackage{amssymb,amsmath}
\usepackage{color}
\usepackage{natbib}
\begin{document}


\title{Neutron scattering evidence for isolated spin-1/2 ladders in (C$_5$D$_{12}$N)$_2$CuBr$_4$}


\author{A.~T.~Savici}
\email[]{saviciat@ornl.gov}
\affiliation{Department of Physics and Astronomy, Johns Hopkins University, Baltimore, MD 21218}
\author{G.~E.~Granroth}
\affiliation{Neutron Scattering Sciences Division, Oak Ridge National Laboratory, Oak Ridge, TN
37831-6477}
\author{C.~L.~Broholm}
\affiliation{Department of Physics and Astronomy, Johns Hopkins University, Baltimore, MD 21218}
\affiliation{NCNR, National Institute of Standards and Technology, Gaithersburg, MD 20899-6100}
\author{D.~M.~Pajerowski}
\affiliation{Department of Physics and NHMFL, University of Florida, Gainesville, FL 32611-8440}
\author{C.~M.~Brown}
\affiliation{NCNR, National Institute of Standards and Technology, Gaithersburg, MD 20899-6100}
\author{D.~R.~Talham}
\affiliation{Department of Chemistry, University of Florida, Gainesville, FL 32611-7200}
\author{M.~W.~Meisel}
\affiliation{Department of Physics and NHMFL, University of Florida, Gainesville, FL 32611-8440}
\author{K.~P.~Schmidt}
\affiliation{Lehrstuhl f\"ur theoretische Physik I, Otto-Hahn-Str.~4, D-44221
Dortmund, Germany}
\author{G.~S.~Uhrig\footnote{On leave at the University of New South Wales, Sydney, Australia.}}
\affiliation{Lehrstuhl f\"ur theoretische Physik I, Otto-Hahn-Str.~4, D-44221
Dortmund, Germany}
\author{S.~E.~Nagler}
\affiliation{Neutron Scattering Sciences Division, Oak Ridge National Laboratory, Oak Ridge, TN 37831-6477}
\date{\today}

\begin{abstract}
Inelastic neutron scattering  was used to determine the spin Hamiltonian for the singlet ground state system of fully deuterated BPCB, (C$_{5}$D$_{12}$N)$_{2}$CuBr$_{4}$. A 2-leg spin-1/2 ladder model, with
$J_\bot = (1.09 \pm 0.01)$~meV and $J_\| = (0.296 \pm 0.005)$~meV, accurately describes the data. The experimental limit on the effective inter-ladder exchange constant is $|J_{\rm int}^{\rm eff}|\lesssim 0.006$~meV, and the limit on total diagonal, intra-ladder exchange is $|J_F+J_{F'}|\leq 0.1$ meV. Including the effects of copper to bromide covalent spin transfer on the magnetic form-factor, the experimental ratios of intra-ladder bond energies are consistent with the predictions of continuous unitary transformation.

\end{abstract}


\maketitle

\section{Introduction}
Low dimensional quantum magnets, with integer spin per unit cell, frequently exhibit a macroscopic singlet ground state. While impervious to weak fields, there is a critical field beyond which an extended critical phase can exist, and it is of considerable interest to explore the spin dynamics at the quantum critical point and within the putative critical phase.

The $n$-leg antiferromagnetic (AFM) spin ladder consists of $n$ parallel chains of magnetic moments with AFM exchange interactions along the chains ($J_\|$) and between neighboring chains ($J_\bot$).
In keeping with the Lieb-Schultz-Mattis theorem,\cite{Lieb61,Affleck88} the excitation spectrum has a gap for $n$ even and is gapless for
$n$ odd.\cite{Rice93,Gopalan94,Dagotto99}  The $n=2$, $S=1/2$ case is of particular interest
because such ladders may form dynamically in copper oxide superconductors and play a role in
suppressing magnetism in favor of superconductivity.  The Hamiltonian for the simple two-leg spin-ladder
is given by the first two terms in
\begin{eqnarray}
\label{hamiltonian}
\nonumber {\cal H} &=& J_\|\sum_{j,l=1,2}\textbf{S}_{j,l}\cdot\textbf{S}_{j+1,l}
+ J_\bot\sum_{j}\textbf{S}_{j,1}\cdot\textbf{S}_{j,2}\\
\nonumber& &+ J_{F}\sum_{j}\textbf{S}_{j,1}\cdot\textbf{S}_{j+1,2}+ J_{F'}\sum_{j}\textbf{S}_{j,1}\cdot\textbf{S}_{j-1,2}\\
\nonumber& &+ J'\sum_{j,m,n}\textbf{S}_{j,m}\cdot\textbf{S}_{j,n}
+ J''\sum_{j,m,n}\textbf{S}_{j,m}\cdot\textbf{S}_{j+1,n}\\
& &-g \mu_B H \sum_{j,l=1,2}\textbf{S}_{j,l} \;\;,
\end{eqnarray}
where $l=1,2$ indexes each of the two chains and $j$ is the rung index. However, for BPCB (see Fig.~\ref{structure}(a)) additional
interactions are possible. The third and fourth terms are
frustrating diagonal intra-ladder exchange interactions, $J_F$ and $J_{F'}$, between spins in neighboring chains
and rungs.  The next two terms describe two possible inter-ladder interactions, $J'$ and $J''$, where
$m$ and $n$ denote adjacent chains in different ladders. We define $J_{\rm int}^{\rm eff}=J'-J''$ to discuss the effective interaction between adjacent ladders. The last term in the Hamiltonian is the Zeeman
term associated with an applied magnetic field, $H$. It is included here in anticipation of future high field experiments, but for this work, $H = 0$.

Consider the ideal 1D system ($J',J'' = 0$). If $J_\|=0$ and one of the frustrating exchanges is zero, ${\cal H}$ describes an alternating
spin chain, with the physics controlled by $\beta=J_F/J_\bot$. Another extreme, $J_F=J_{F'}=0$, is the
ideal spin ladder, where the physics is controlled by $\alpha=J_\|/J_\bot$. In the limit
$|\alpha| \rightarrow \infty$, the system is composed of decoupled 1D chains with a gapless
spectrum.\cite{Nagler91,Tennant95,Lake05,Hammar99,Stone03}  Any finite $J_\bot$ produces an isolated
singlet ground state with a gap $\Delta\approx |J_\bot|/2$ (see Ref.~\onlinecite{Dagotto99}).
This general state of affairs persists into the strong coupling limit, $|\alpha| \ll 1$,  where
the ground state is a singlet, separated from the lowest lying triplet of excited states by an
energy gap $\Delta \approx J_\bot - J_\|$.\cite{Barnes93,Reigrotzki94}

Given the extensive theoretical and numerical treatments that Eq.~(\ref{hamiltonian}) has received,\cite{batchelor07}
it is of great interest to identify materials containing spin-1/2 ladders that can be
driven to quantum criticality through the application of a magnetic field. Excluding the system
described in this paper, spin ladder materials known thus far either have energy scales that are
too large to be affected by an applied field\cite{Honda97} or have significant inter-ladder interactions
that induce N\'{e}el order above the critical field.\cite{Masuda06}  Diagonal\cite{Garlea07,Masuda06}
and cyclic\cite{Notbohm07} exchanges, along with the possible coexistence with other magnetic
systems,\cite{Eccleston98,Notbohm07,Lemmens03,Hess07} represent additional challenges that are of
interest in their own right but that may complicate analysis of the quantum critical phenomena.

Here we report experimental evidence that fully deuterated BPCB, (C$_{5}$D$_{12}$N)$_{2}$CuBr$_{4}$, is a
nearly ideal realization of an assembly of non-interacting spin-1/2 ladders. We provide an accurate determination
of the relevant exchange constants, and quantify the low temperature exchange bond energies, in the absence of any applied magnetic field. These values can be used for describing the field induced quantum critical state in other experiments.

\section{Experimental details}
\begin{figure}[!t]
\begin{center}
\includegraphics[width=3.1in,angle=0]{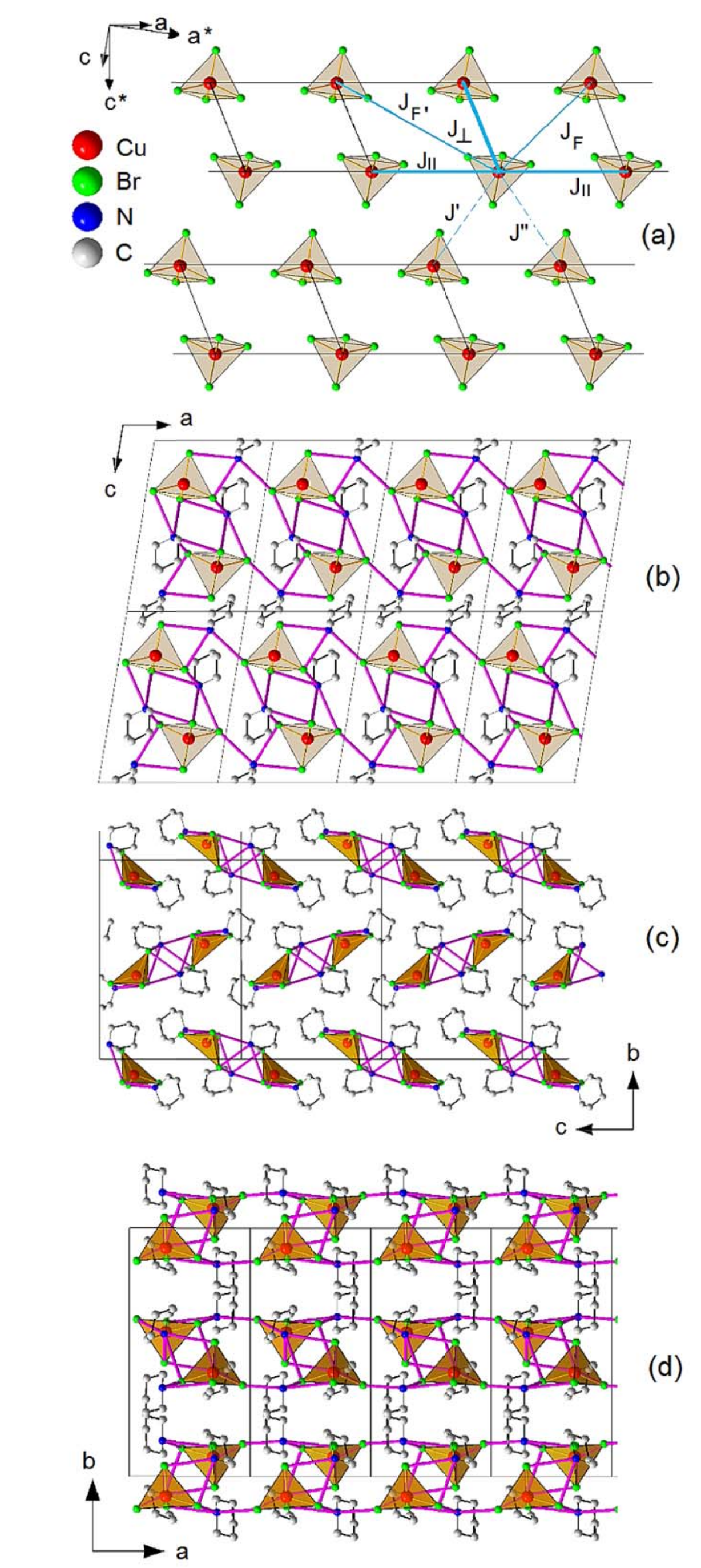}%
\caption{\label{structure} (Color online) Various projections of the previously determined structure
of BPCB.\cite{Patyal90} (a) Exchange interactions entering the Hamiltonian in Eq.~(\ref{hamiltonian}).
Black lines indicate the legs and rungs of the ladder.  (b-d) BPCB crystal  structure, except hydrogen,
projected along each of the crystallographic axes.}
\end{center}
\end{figure}
We used inelastic neutron scattering to probe magnetic excitations in fully deuterated
bis(piperidinium)tetrabromocuprate(II), commonly referred to as BPCB, (C$_5$D$_{12}$N)$_2$CuBr$_4$.
BPCB is monoclinic (space group $P2_1/c$), with room temperature lattice parameters $a=8.49$~\AA, $b=17.22$~\AA, $c=12.38$~\AA, and $\beta=99.3^{\circ}$.\cite{Patyal90} Throughout we shall denote wave vector transfer in the corresponding reciprocal lattice ${\bf q}(hkl)=h{\bf a}^*+k{\bf b}^*+l{\bf c}^*$. The Cu$^{2+}$ ions form ladders,
as shown in Fig.~\ref{structure}. The legs of the ladders run along ${\bf a}$, and the rungs are
nearly along ${\bf c}^*$, with a small tilt of $24^{\circ}$ above the $a-c$ plane.\cite{Patyal90, Watson01}

High field magnetization measurements ($0<H < 30$~T) were performed by Watson
\textit{et al.},\cite{Watson01} revealing a lower critical field of $H_{c1}=6.6$~T,
an upper critical field of $H_{c2}=14.6$~T, and an inflection point at half the saturation
magnetization. Through careful comparison of bulk thermo-magnetic data to various models,
BPCB was identified as a two-leg spin ladder in the strong coupling limit with $J_\bot/k_B=13.3$~K
(1.15~meV) and $J_{\|}/k_B=3.8$~K (0.33~meV).  The analysis indicated that BPCB possesses an isolated
singlet ground state for $H<H_{c1}$ and forms a gapless Luttinger spin liquid for $H_{c1}<H<H_{c2}$.

Inelastic neutron scattering is a sensitive probe of atomic scale correlations and interactions in
singlet ground state systems. Using this technique, we find that BPCB is highly one-dimensional
($|J_{\rm int}^{\rm eff}/J_\bot| \lesssim 5 \times 10^{-3}$), making it an excellent candidate for future exploration of the high field
critical phase. The findings extend, and are consistent with, previous experimental results on the
spin Hamiltonian for BPCB.

Using 99.9\% deuterated starting materials, the sample used for our neutron scattering measurements was made
by the same process and team of scientists as described in Ref.~\onlinecite{Watson01}. The sample consists of
five deuterated single crystals, with a total mass of 3~g and co-aligned within one degree, for scattering
in the $(h0l)$ reciprocal lattice plane. The measurements were performed at the NIST Center for Neutron
Research using the time-of-flight Disk Chopper Spectrometer (DCS).\cite{Copley03}  The chopper cascade
was phased to provide an incident wavelength $\lambda=5$~\AA\ and energy resolution $\Delta E \sim 0.1$~meV.
The sample was cooled in a liquid helium cryostat to $T = (1.4 \pm 0.1)$~K.  The non-magnetic background
was measured at $T=25$~K, where magnetic scattering is widely distributed in energy and momentum.
To a good approximation, the high temperature scattering can be treated as independent of sample
orientation. This background measurement was subtracted from the 1.4~K data.\cite{Tsujii02}  The DAVE software package was used
to perform the initial analysis, to compute the energy resolution, and to extract the data required
for advanced processing.\cite{DAVE}

\begin{figure}[!t]
\begin{center}
\includegraphics[width=3.3in,angle=0]{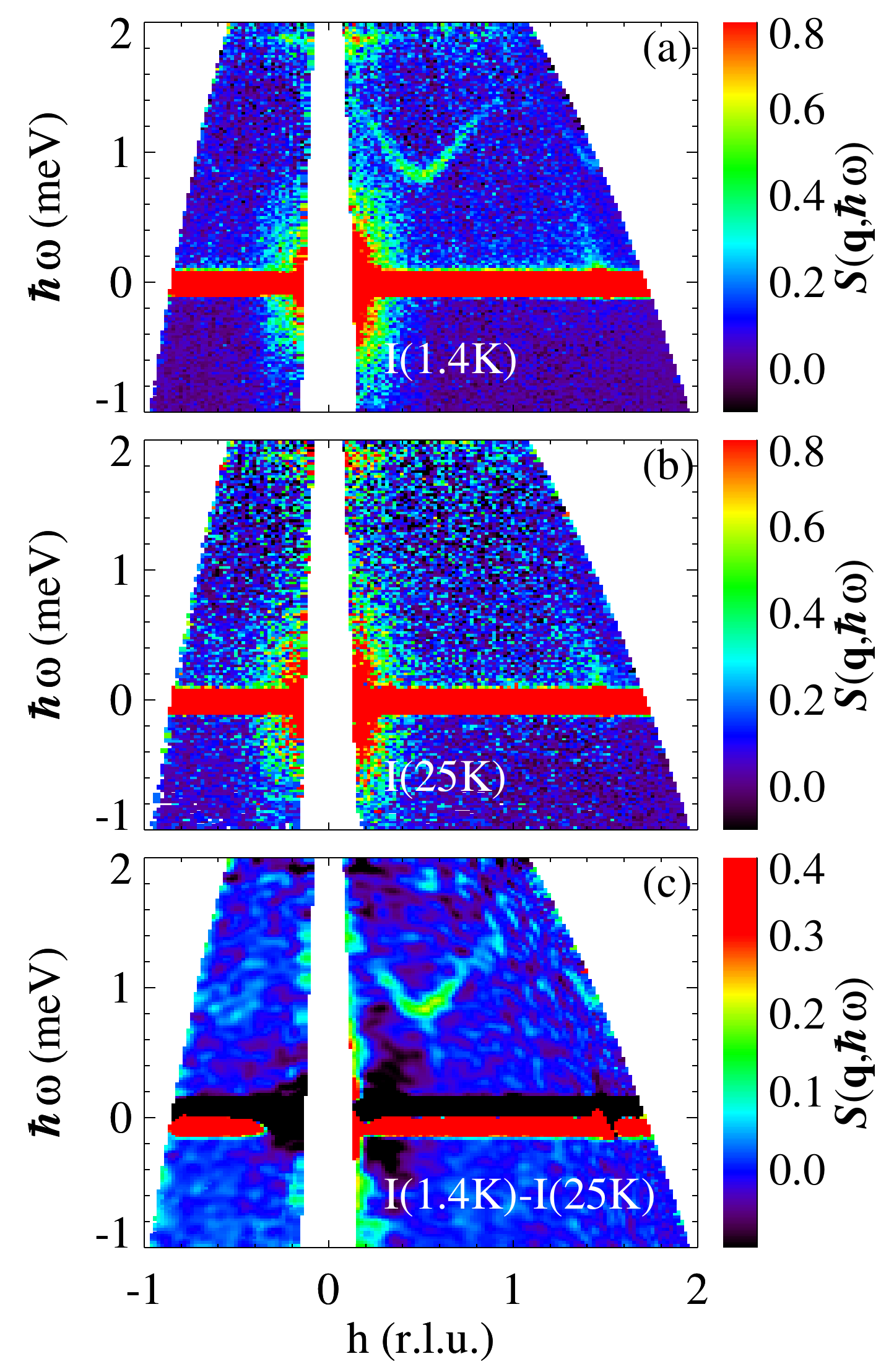}%
\caption{\label{data} (Color online) False color image of raw $\lambda_i=5$ \AA\
neutron scattering data measured with ${\bf c}^*\|{\bf k}_i$. Intensity is integrated in the $l$
direction. (a) $T=1.4$~K. (b) $T=25$~K, used as background. (c) Intensity at $T = 1.4$~K after background
subtraction. For presentation purposes only, the data in (c) were smoothed as described in the text
(see Sec.~\ref{sectionB}).}
\end{center}
\end{figure}
\begin{figure}[!t]
\begin{center}
\includegraphics[width=3.3in,angle=0]{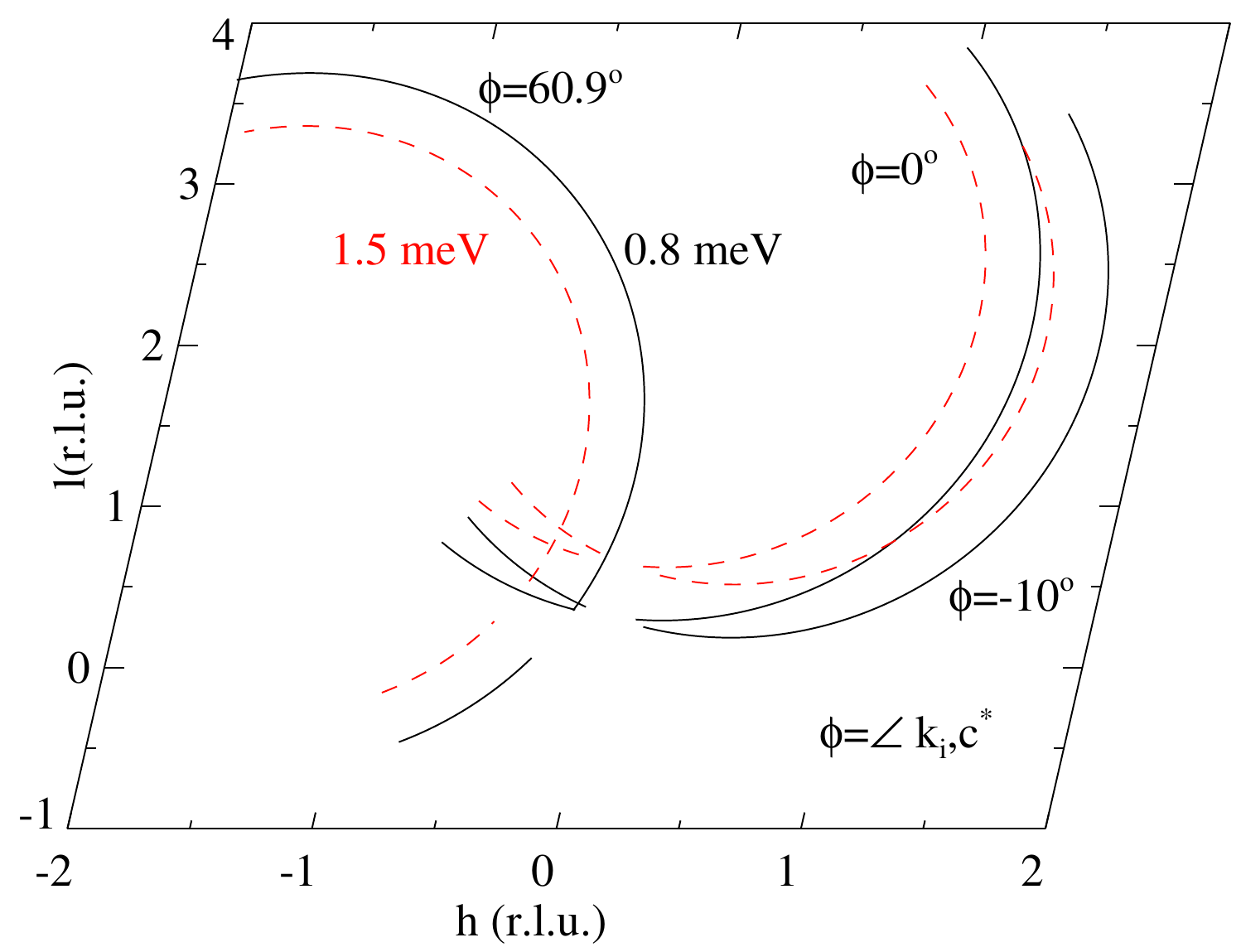}%
\caption{\label{traj} (Color online) Trajectories in $(h,0,l)$ plane for the three sample orientations
measured at $\hbar\omega = 0.8$~meV and 1.5~meV, with
$\lambda_i = 5$~\AA.}
\end{center}
\end{figure}
An example of raw data measured with the ${\bf c}^*$ axis parallel to ${\bf k}_i$ is shown in
Fig.~\ref{data}(a). An integration over the $k$ and $l$ directions is performed to generate this figure.
Such a procedure is commonly used for analyzing data acquired on low dimensional systems using
chopper spectrometers.\cite{Xu00} However, we will use our measurements to quantify inter-ladder exchange.
For this purpose, all components of momentum transfer in the horizontal scattering plane were used in
the subsequent analysis. The trajectories in the ($h,0,l$) plane are shown in Fig.~\ref{traj} for
energy transfers of 0.8 meV and 1.5 meV, and for all sample orientations used
($\phi \equiv \angle {\bf k}_i$, ${\bf c}^* = 0^\circ$ , $-10^\circ$, and $+60.9^\circ$).  The trajectories sample the ($h,0,l$)
plane sufficiently to test for the presence of inter-ladder dispersion.

\section{Analysis and discussion}
\subsection{Exchange paths}

As a first step towards a model spin Hamiltonian, we shall discuss the structure and chemical bonding in
BPCB. A review of the magnetic exchange interactions in a wide-range of tetrabromocuprates\cite{Turnbull05} provided guidance on possible exchange paths.
While these considerations are not rigorous, they can provide a reference against which to compare the
experimental results.  It was proposed\cite{Patyal90,Watson01} that the rung interaction ($J_\bot$) is
associated with overlap of Br$^-$ adjacent to copper sites, while the exchange interactions along the legs
of the ladder ($J_{\|}$) are mediated by a combination of hydrogen bonds and non-overlapping Br$^-$ orbitals,
and therefore should be weaker. However, Fig.~\ref{structure} shows that the Br-Br distances associated
with inter- and intra- chain interaction are in fact quite similar. Furthermore, these distances are
$4-5$ \AA, which is more than twice the covalent radius of bromine. This observation indicates that magnetic
interactions in BPCB are mediated by intervening hydrogen, as becomes increasingly clear when piperidinium radicals
are included in the picture, Fig.~\ref{structure}(b-d). Excess hydrogen is located around the nitrogen sites
in the piperidinium rings, and these are found aligned with the nearest approach of bromine atoms,
associated with neighboring Cu$^{2+}$ ions. Two piperidinium groups are involved in producing $J_\bot$,
and only one for $J_\|$. The greater number of Br-H bridges for rung over leg interactions leads to
an expectation of dominant rung exchange.

Any frustrating diagonal interaction ($J_F$ or $J_{F'}$) would involve traversing the piperidinium molecule. Note that a
large $|J_F|>J_\|$ and $J_{F'}=0$ would result in an alternating chain, as opposed to a ladder. This situation is found,
for example, in MCCL.\cite{Stone07}
Furthermore, $J_F$ and $J_{F^{\prime}}$ being associated with different
bond lengths of 8.96~\AA\  and 12.64 \AA, respectively, suggests $J_F$
dominates.

The strongest inter-ladder interaction, $J'$, is expected between ladders
separated by {\bf c}, and it is
mediated by hydrogen bonding through the same piperidinium molecule involved in the leg exchange. In addition, a $J''$ inter-ladder exchange interaction
is possible between atoms in ladders separated by {\bf c} + {\bf a}. In conjunction with $J'$, a finite
$J''$ interaction might
produce frustration, giving a small $J_{\rm int}^{\rm eff}$, and reduce inter-ladder dispersion. Any inter-ladder exchange in the {\bf b} direction
would involve a longer path, through two piperidinium molecules, and is therefore expected to be weak.

\subsection{Global Fitting based on Single Mode Approximation}
\label{sectionB}
The single mode approximation\cite{Hohenberg74} generally provides an excellent description of the
dynamic spin correlation function for gapped quantum magnets. The assumption that all spectral weight
resides in a dispersive ``triplon'', combined with the first moment sum-rule, leads to the following
expression for ${\cal S}({\bf q},\hbar \omega)$:

\begin{equation}
\label{SMA}
{\cal S}({\bf q},\hbar \omega)= -\frac{1}{3} \frac{\delta(\hbar \omega-E_{\bf q})}{E_{\bf q}} \sum_{\bf d}
\varepsilon_{\bf d} [1-\cos(\textbf{q}\cdot\textbf{d})]\;\;\;,
\end{equation}
where $E_{\bf q}$ is the triplon dispersion relation and
$\varepsilon_{\bf d} = J_{\bf d} \langle\textbf{S}_0\cdot\textbf{S}_\textbf{d}\rangle$ are the so-called
bond energies, which sum to the ground state energy for $T=0$. The summation is over spin pairs with
finite exchange interactions. For the Hamiltonian in Eq.~(\ref{hamiltonian}), in each unit cell there is one $J_\bot$
rung spin-pair term, two $J_\|$ terms, one $J_F$ term, one inter-ladder $J'$ term, and one inter-ladder $J''$ term, as shown in Fig.~\ref{structure}(a). The dispersion relation, $E_{\bf q}$, is a function of $J_\bot$,
$\alpha=J_\|/J_\bot$, $\beta=J_F/J_\bot$, and $\gamma=J_{\rm int}^{\rm eff}/J_\bot$. The neutron scattering intensity is
obtained by multiplying ${\cal S}({\bf q},\hbar \omega)$ by the square of the Cu$^{2+}$ magnetic form factor\cite{Brown03}
and convoluting the result with the instrumental resolution.

The raw data binning results in an effective wave-vector resolution approximatively equal to the pixel size.  The energy resolution was wider than the pixel size, and it is described by a Gaussian,
with the width depending on the configuration of the instrument and on the incident and scattered
neutron energies. All subsequent analysis focuses on scattering data in the 0.7~meV to 1.7~meV energy range.
Global fits to the subtracted 25~K background were performed with all three sample orientations
simultaneously. Any residual background was treated as momentum and energy independent.
To generate false color images of the treated data for Fig.~\ref{data} (and Figs. \ref{fits} and \ref{dimers}, which will appear
subsequently), a 3D Gaussian smoothing was used
with full-widths at half-maximum of 0.03 meV and 0.03~r.l.u.~along the $h$ and $l$ directions.

\subsubsection{Triplon dispersion}
\begin{figure}[!t]
\begin{center}
\includegraphics[width=3.3in,angle=0]{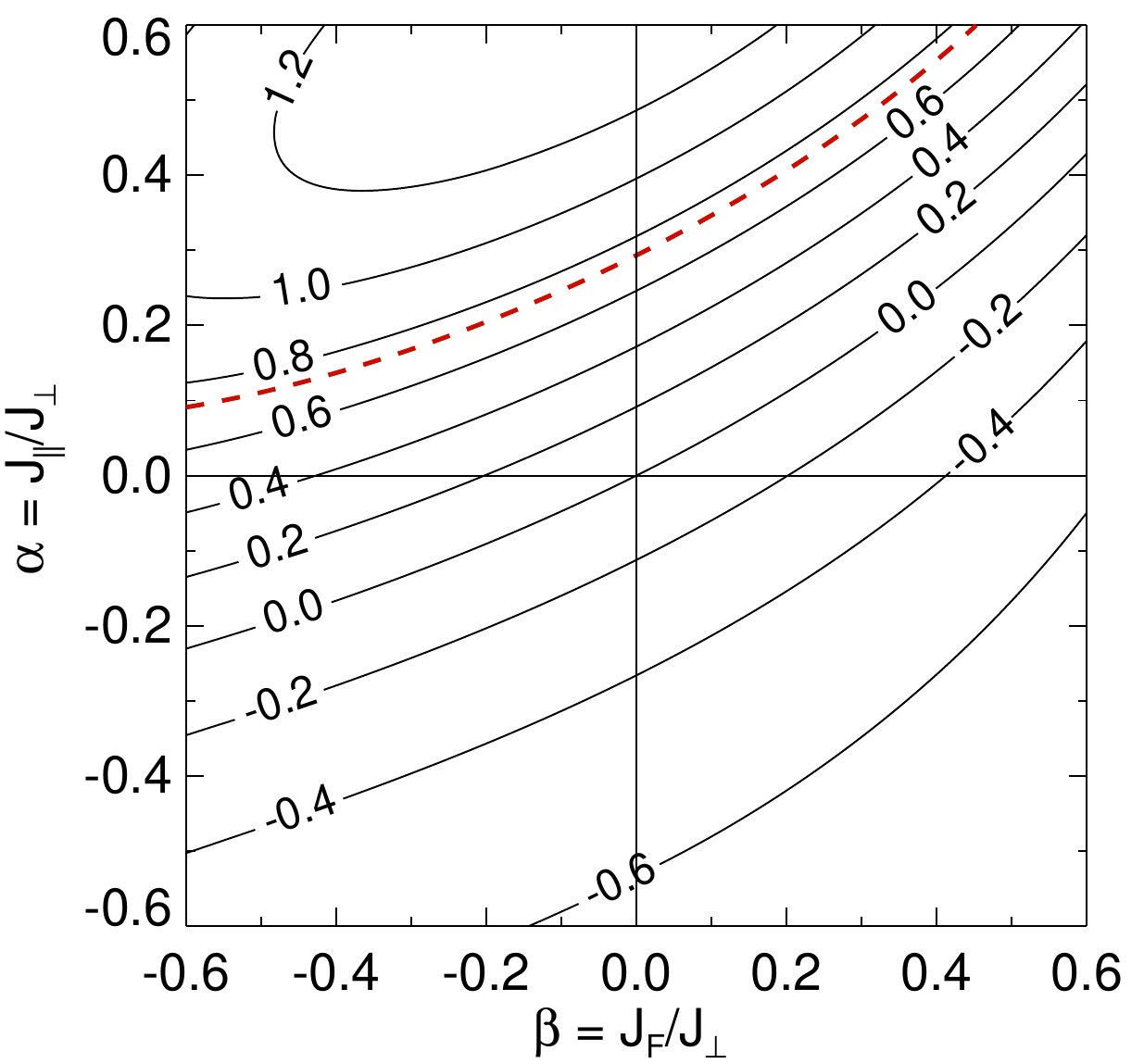}%
\caption{\label{Widgap} (color online) Bandwidth over bandgap ($W/\Delta$) ratio for spin excitations
versus normalized intra-ladder exchange interactions, computed using Eqs.~(\ref{omega},\ref{muller}). The experimental
result of $W/\Delta = 0.73$ for BPCB constrains intra-ladder exchange interactions
to lie on the dashed red line.}
\end{center}
\end{figure}
\begin{figure*}
\begin{center}
\includegraphics[height=4in,angle=0]{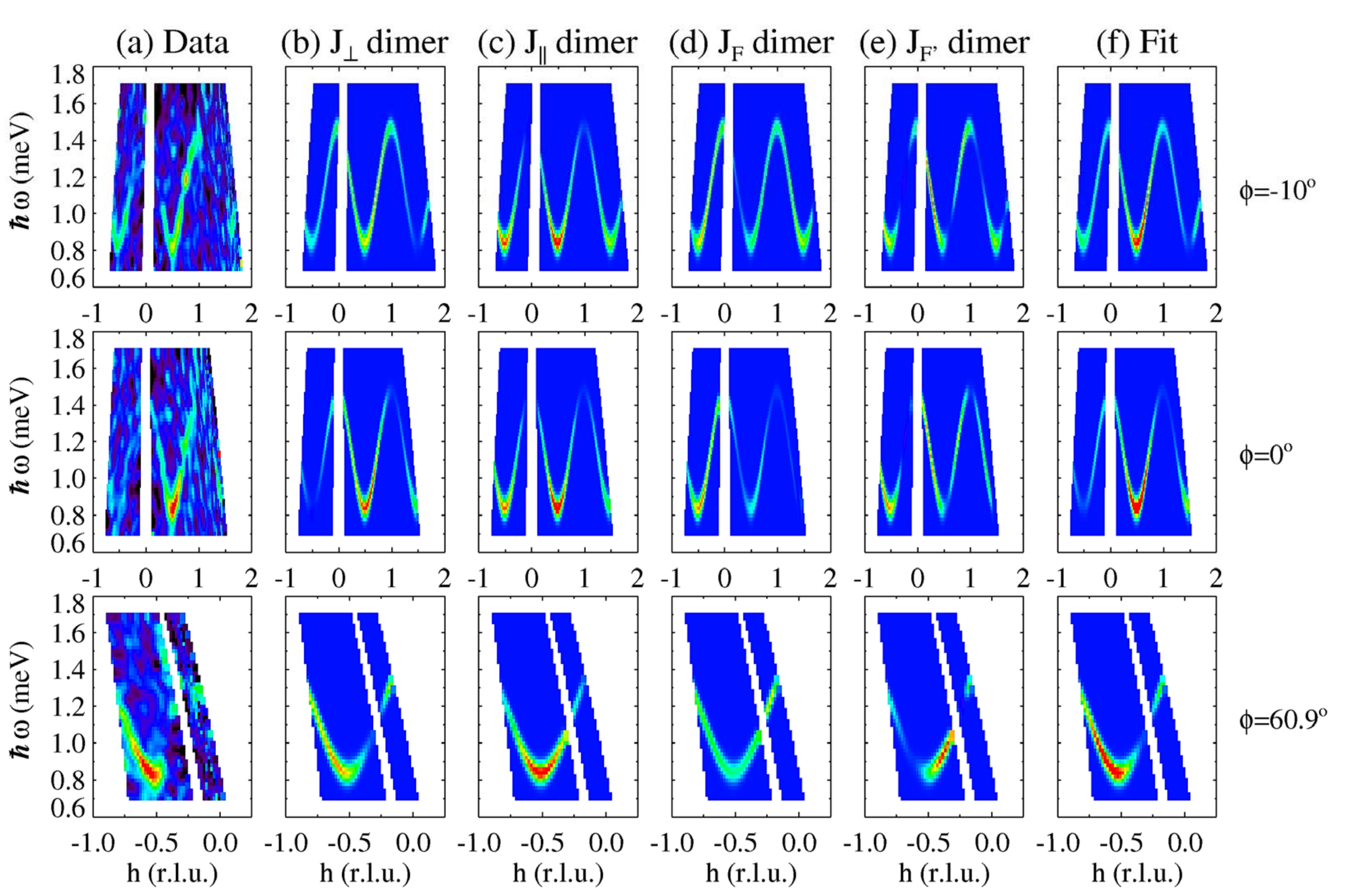}%
\caption{\label{dimers} (color online) False color images of neutron scattering intensity
for three sample orientations (rows): (column a) smoothed (see text), background-subtracted,
data; (columns b-e) simulations with the phenomenological dispersion relation, Eq. (\ref{simpledisp}),
but only one bond energy (see top) modulating the intensity for each column; (column f) fit
including all bond energies. For each sample orientation, select sections of data with unique values of
$h$ and $l$ are shown. The color scale is the same one used in
Fig. 2c. }
\end{center}
\end{figure*}

To quantify the magnetic interactions in BPCB, Eq.~(\ref{hamiltonian}), we examine the triplon
dispersion relation.  Figure~\ref{data} suggests that a lowest order approximation to the dispersion
is a periodic lattice sinusoidal function of the form
\begin{equation}
\label{simpledisp}
E_{\bf q}=\Delta+\frac{W}{2}(1+\cos(2\pi h))+A_{c*}\cos(2\pi l) \;\;\;,
\end{equation}
where $\Delta$ denotes the gap in the absence of inter-ladder dispersion, $W$ is the intra-ladder bandwidth, and $A_{c*}$ is the amplitude
of dispersion along ${\bf c}^*$, resulting from $J_{\rm int}^{\rm eff}$. We found that this last term is zero,
within experimental uncertainty, but a more detailed analysis will be employed later to establish
an experimental limit. The global fit (Fig.~\ref{dimers}(f)) yields a bandwidth
$W = (0.62 \pm 0.03)$~meV, and a spin gap $\Delta = (0.85 \pm 0.01)$~meV. The latter value can be
compared with $\Delta=0.82$~meV, obtained from magnetization measurements \cite{Watson01} on a hydrogenous
powder sample, where $H_{c1}=6.6$ T and $\langle g \rangle =2.13$. The structure factor indicates that the
$\varepsilon_\bot$ bond energy is dominant. The next step towards a spin Hamiltonian for BPCB is to
relate the phenomenological parameters characterizing the dispersion relation to exchange constants.

For a strictly one-dimensional model ($J',J''=0$), we can use perturbative expressions for the dispersion
relation to extract exchange constants from the data.
Contributions to dispersion from $J_F$ and $J_{F'}$ cannot be distinguished when both are small, so we define $\bar{J_F}=(J_F+J_{F'})/2$.
When both $\alpha=J_\|/J_\bot$ and $\beta=\bar{J_F}/J_\bot$
are present, the model is a ladder with frustrating diagonal exchange, or equivalently, an alternating
spin chain with next nearest neighbor exchange. In either case, the dispersion is given by\cite{Muller00}
\begin{equation}
\label{omega}
\frac{E(h)}{J_\bot}=\sum_{m=0}^\infty a_m(\alpha,\beta)\cos(2\pi m h)
\end{equation}
with
\begin{eqnarray}
\label{muller}
\nonumber a_0 & = & 1-\beta^2 \frac{1+\alpha}{4}+ \frac{3}{8} (\alpha-\frac{\beta}{2})^2
(2+\alpha-\frac{\beta}{2})+\ldots\\
a_1 & = & \alpha -\frac{\beta}{2}-\beta^2 \frac{1+\alpha}{4}- \frac{(\alpha-\frac{\beta}{2})^3}{4}+
\ldots\\
\nonumber a_2 & = & -\frac{1}{4}(\alpha-\frac{\beta}{2})^2(1+\alpha+\frac{\beta}{2})+\ldots\\
\nonumber &\ldots \;\;\;.
\end{eqnarray}
\begin{figure}[!b]
\begin{center}
\includegraphics[width=3.3in,angle=0]{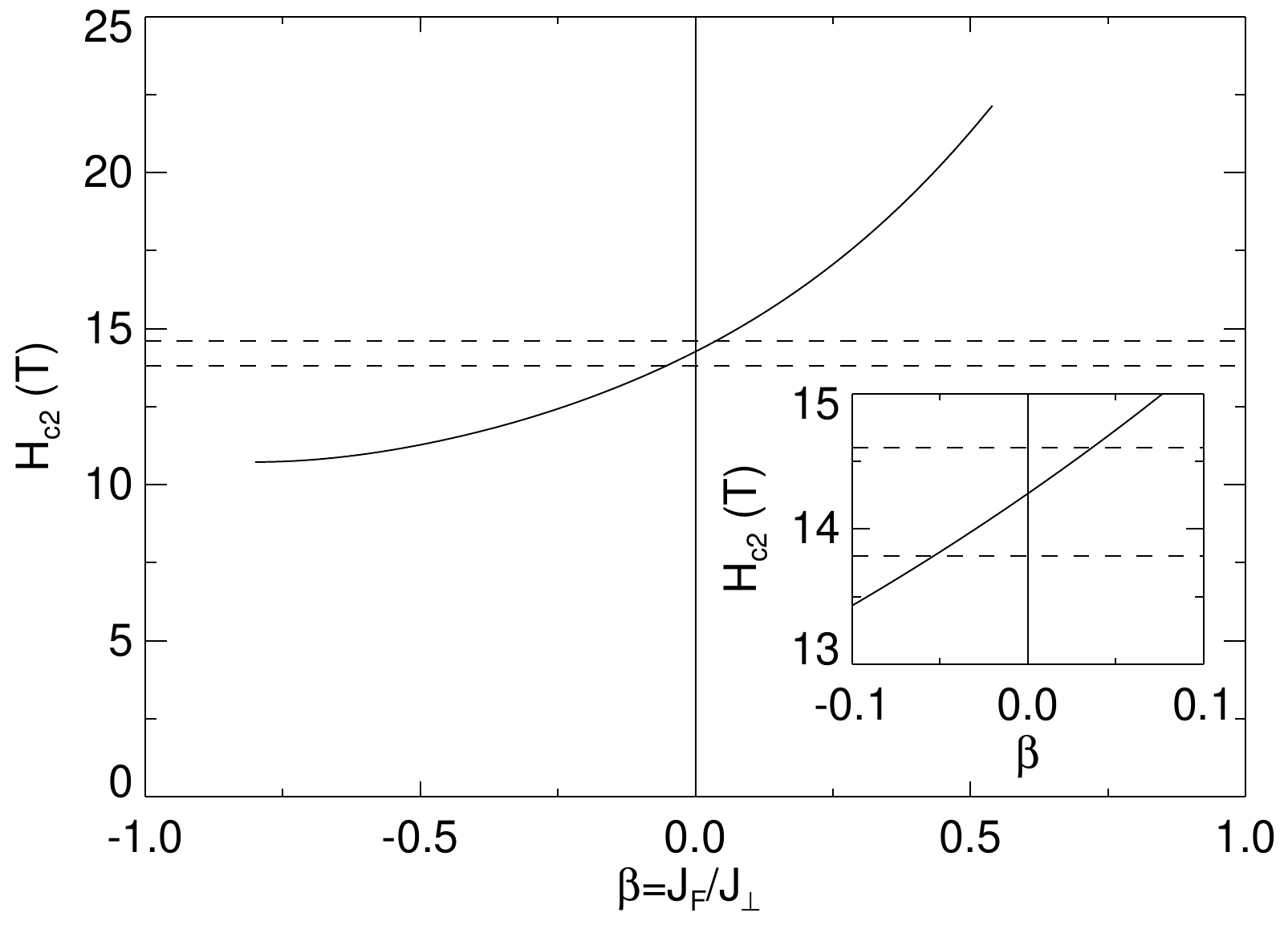}%
\caption{\label{Hc2} $H_{c2}$ as a function of $J_F/J_\bot$, for $W/\Delta=0.73$. The dashed lines
correspond to $H_{c2} = 13.8$~T and $H_{c2} = 14.6$~T, as reported in different
references.\cite{Anfuso08,Watson01} Several other measurements of the upper critical field
lie in this range.\cite{Klanjsek08,Lorenz08,ruegg08,Thielemann08}  Detail shown in the inset
indicates that $|J_F/J_\bot| \leq 0.05$, so $|J_F/k_B| \leq 0.4$~K.}
\end{center}
\end{figure}
\begin{figure}[!t]
\begin{center}
\includegraphics[width=3.3in,angle=0]{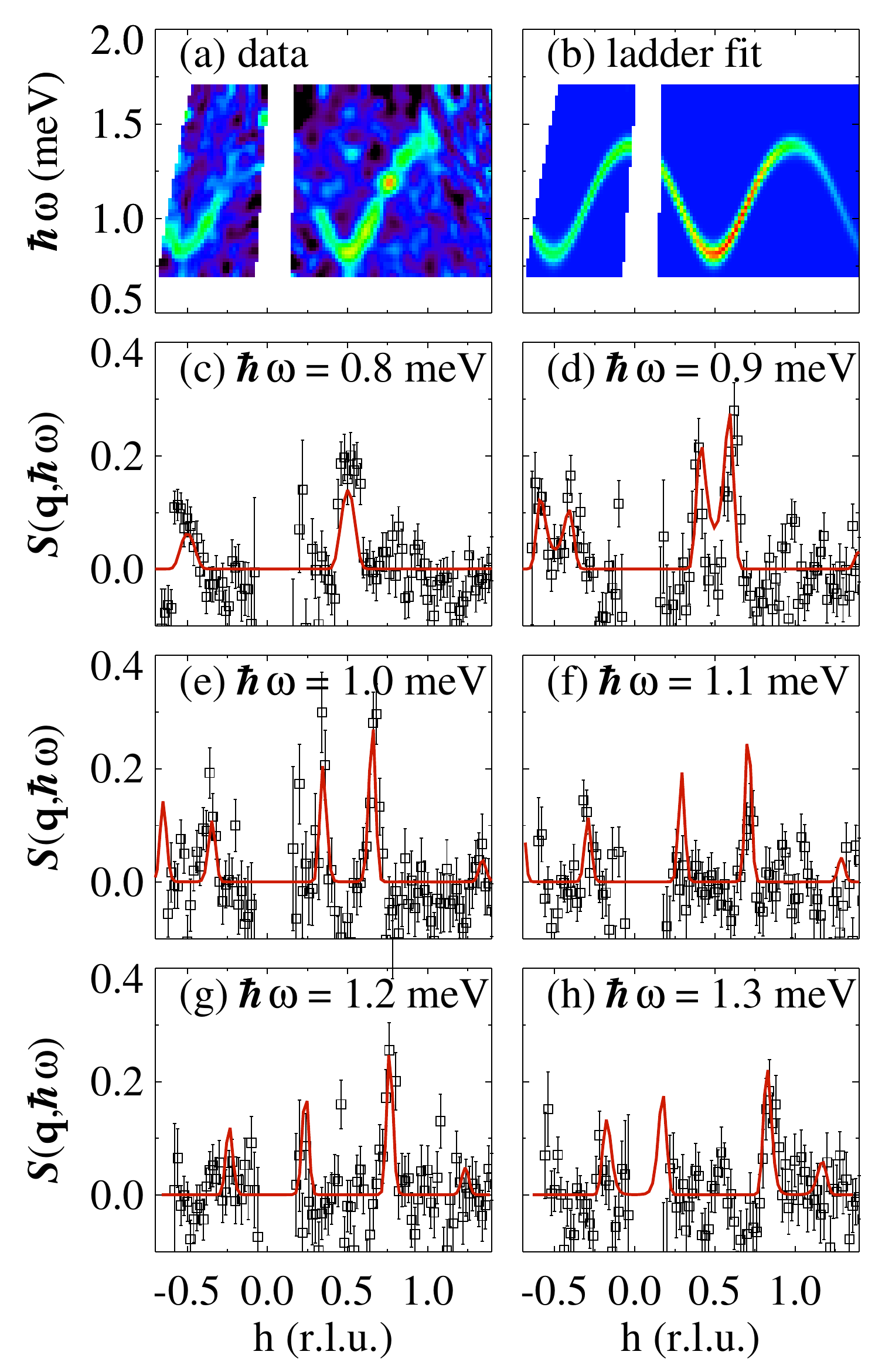}%
\caption{\label{fits} (Color online) Global fit to a spin ladder model.
(a) color map of smoothed (see text), background-subtracted data, with $\phi=\angle {\bf k}_i$,
${\bf c}^*=-10^\circ$.The color scale is the same one used in
Fig. 2c; (b) color map of the corresponding fit; (c)-(h) constant energy cuts,
with an energy window of 0.1 meV. The solid lines represent the global fit to the data when
including inter-ladder coupling effects.}
\end{center}
\end{figure}

To account for inter-ladder exchange, in the first approximation, we add $\gamma \cos(2 \pi l)$ to Eq.~(\ref{omega}). For stronger coupling, or different inter-ladder exchange paths,\cite{Thielemann09} one expects more complicated $l$ dependence, and possible cross terms involving both $h$ and $l$ dependence.\cite{Matsumoto04}
In the absence of $J''$ exchange, $\gamma = J'/J_\bot$.  If $J''$ is present and frustrates $J'$,
$\gamma = J_{\rm int}^{\rm eff}/J_\perp = (J'-J'')/J_\perp$. In principle there should also be a {\bf q}-independent term associated
with $J_{\rm int}^{\rm eff}$. This effect is, however, sufficiently small to be neglected.

With $J_\bot$ dominant, there are two limiting cases: $J_\|=0$ or $J_F=0$.  When $J_\|=0$, the system is
an alternating spin chain, with dispersion controlled by $\beta=J_F/J_\bot$.\cite{Barnes03}
From the $a_1$ term, we note that BPCB can be described only by a negative $\beta$, so $J_F$ would have
to be ferromagnetic, corresponding to a FM/AFM alternating spin chain. However, calculations show that
one cannot achieve the experimental bandwidth over bandgap ratio, $W/\Delta=0.73$, in the small $\beta$
limit where Eq.~(\ref{muller}) is valid.  In addition, a strong ferromagnetic interaction is incompatible
with magnetization measurements.

A contour map of the $W/\Delta$ ratio as a function of $\alpha$ and $\beta$ is shown in Fig.~\ref{Widgap}.
The dashed line corresponds to all ($\alpha$, $\beta$) pairs that are consistent with $W/\Delta=0.73$.
For all points on this line, with $|\beta| \leq 0.5$, $a_0$ varies by less than 3\%. Hence, any set of
($\alpha$, $\beta$) on this line accurately describes the observed dispersion relation, with the value of
$J_\bot$ within 3\% of 1.09 meV. An additional constraint is therefore required to uniquely determine
($\alpha$, $\beta$). This is provided by previous measurements of the upper critical
field $H_{c2}$.\cite{Anfuso08,Watson01,Klanjsek08,Lorenz08,ruegg08,Thielemann08}
For a spin ladder with frustrating diagonal exchange, the upper critical field is
\begin{equation}
g \mu_B H_{c2}\;=\;J_\bot\,+\,2 J_\| \;\;\;,
\end{equation}
where $H_{c2}$ is independent of $J_F$, if $J_F < J_\bot$ and $J_F < 2 J_\|$.\cite{Mila98} Figure~\ref{Hc2} shows $H_{c2}$
versus $J_F/J_\bot$ for values ($\alpha$, $\beta$) along the dashed line in Fig.~\ref{Widgap}.
The values found experimentally vary between 13.8 T (Ref.~\onlinecite{Klanjsek08}) and
14.6 T (Ref.~\onlinecite{Watson01}),
as indicated by dashed lines in the figure and in the insert. From this analysis, we conclude that
$|J_F/J_\bot| \leq 0.05$ or $|J_F/k_B| \leq 0.4$~K. As an alternative method, we included
the values of the upper critical field as a constraint in the fit to neutron scattering data and this yields $\beta=J_F/J_\bot=-0.02\pm0.10$.
These results are consistent with the upper limit reported by Watson {\it et al.},\cite{Watson01}
based on magnetization measurements performed at 700~mK. On the basis of this tight limit, the frustrating exchange and
the corresponding bond energy were neglected in the subsequent analysis of scattering data.

When $J_F$ is neglected, Eq.~(\ref{muller}) describes the dispersion for an ideal
ladder.\cite{Barnes03,Reigrotzki94}  A global fit to the scattering data yields
$J_\bot = (1.09\pm0.01)$~meV, $J_\| = (0.296\pm0.005)$~meV, and $\gamma = (0.002\pm 0.006)$. The quoted error bars reflect systematic error estimated as 10\% of the energy resolution. The statistical errors reported by the fitting routine were a factor of 2-3 smaller.

The fitted {\bf q} and $E$ dependent intensity calculated for this model is shown in Fig.~\ref{fits}(b).
A portion of the data, together with the fit, is presented as several constant energy cuts in
Fig.~\ref{fits}(c-e).  The values for $J_\bot$ and $J_\|$ are in excellent agreement with the values
obtained from neutron scattering,\cite{Thielemann08,Thielemann09} magnetization,\cite{Watson01} NMR,\cite{Klanjsek08}
magnetostriction,\cite{Lorenz08,Anfuso08} and specific heat and magnetocaloric effect\cite{ruegg08} measurements.
We note that data has been fitted to an expression valid to third order of
$\alpha$ and $\beta$, Eqs.~(\ref{omega}) and (\ref{muller}).  In principle, much higher order expressions
can be obtained using other methods, including a particle conserving continuous unitary
transformation (CUT)\cite{Knetter00, Knetter03EPJ, Knetter03Jphys} and linked-cluster-expansion
methods.\cite{Hamer03,Oitmaa96,Zheng06}   Given the small values of $\alpha$ and $\beta$, including higher
order terms in the fitted dispersion is unnecessary for BPCB.

\subsubsection{Exchange bond energies}

Within the single mode approximation, Eq.~(\ref{SMA}), the exchange bond energies,
$\varepsilon_{\bf d}=J_{\bf d}\langle\textbf{S}_0\cdot\textbf{S}_\textbf{d}\rangle$,
modulate the neutron scattering intensity periodically in ${\bf q}\cdot{\bf d}$.
A simulation of the intensity pattern, considering each possible exchange path in
isolation, is shown in Fig.~\ref{dimers}, where each row corresponds to a different sample orientation.
The observed $h$ dependence of the intensity (column a) closely resembles that associated with the
$J_\bot$ bond (column b). The $J_\|$ term yields a periodic modulation of intensity that is dissimilar
to the data, and all other terms have intensity minima where the data has maxima.

The fit to Eq.~(\ref{SMA}) (see results in Table \ref{table}) yields an unreasonably large bond energy for the inter-ladder dimer considering the weak interactions.  A possible explanation is that it is not appropriate to use an isotropic spin-only form factor for the Cu$^{2+}$ ion.\cite{Brown03} Hubbard and Marshall showed that covalent bonds strongly affect neutron scattering intensities and magnetic form factors.\cite{Hubbard65} For the particular case of CuBr$_{4}^{2-}$ anion,\cite{Awwadi08}
EPR found,\cite{Chow73} and calculations confirmed,\cite{Yu07} that the electron density is significantly shifted from the copper $d$-orbitals into the $\sigma$ ligand orbitals. In the absence of a calculated magnetic form factor for BPCB similar to the one for cuprate spin chains,\cite{Walters08} we chose to modify the Cu$^{2+}$ ionic form factor by isotropically rescaling its {\bf q} dependence by a factor $r$ to account for the spin density transfer to bromine.

The global fit (Table \ref{table}),
shown in Fig.~\ref{dimers} (column f), yields  $\varepsilon_{\|}/\varepsilon_{\bot}=(0.05\pm0.02)$,
$\varepsilon_{F}/\varepsilon_{\bot}=(0.02\pm0.03)$, and $\varepsilon_{J',J''}/\varepsilon_{\bot}=(\pm0.07\pm0.15)$.
A fit using the bond energies of both diagonal exchanges is not a significant improvement compared to these results, and the bond energy ratios presented above are essentially unchanged.

\begin{table}[b]
\caption{Fit results with an ionic magnetic form factor and with a modified form factor that accounts for covalency effects. For comparison, we show results from other neutron scattering experiments}
\begin{tabular}{c|c|c|c}
  \hline
  & Ionic & Covalent & Other results\\
  &form factor & form factor & \\
  \hline
  $J_\bot$ (meV) & $1.09 \pm 0.01$ & $1.09\pm0.01$ & 1.13$\pm$0.01\cite{Thielemann09}\\
  $J_\|/J_\bot$ & $0.288\pm0.057$ & $0.272\pm 0.002$& 0.252$\pm$0.038\cite{Thielemann09} \\
  $\overline{J_F}/J_\bot$ & $-0.02\pm 0.10$ & $-0.02\pm 0.10$ &-\\
  $J_{int}^{eff}/J_\bot$ & $-0.003 \pm0.003$ & $0.002 \pm 0.006$& 0.007\cite{Thielemann08} \\
  $\varepsilon_{\|}/\varepsilon_{\bot}$ & $0.10 \pm 0.02$ & $0.05\pm0.02$ &-\\
  $\varepsilon_{F}/\varepsilon_{\bot}$ & $-0.02\pm0.04$& $0.02\pm0.03$ &-\\
  $\varepsilon_{J',J''}/\varepsilon_{\bot}$ & $-0.27 \pm 0.04$ & $\pm0.07\pm0.15$ &-\\
  $r$ & 1.00$\pm$ 0.00 & 2.36$\pm$0.13&- \\
  $\chi^2$&1.303&1.293&-\\
  \hline
\end{tabular}
\label{table}
\end{table}

\begin{figure}[!h]
\begin{center}
\includegraphics[width=3.3in,angle=0]{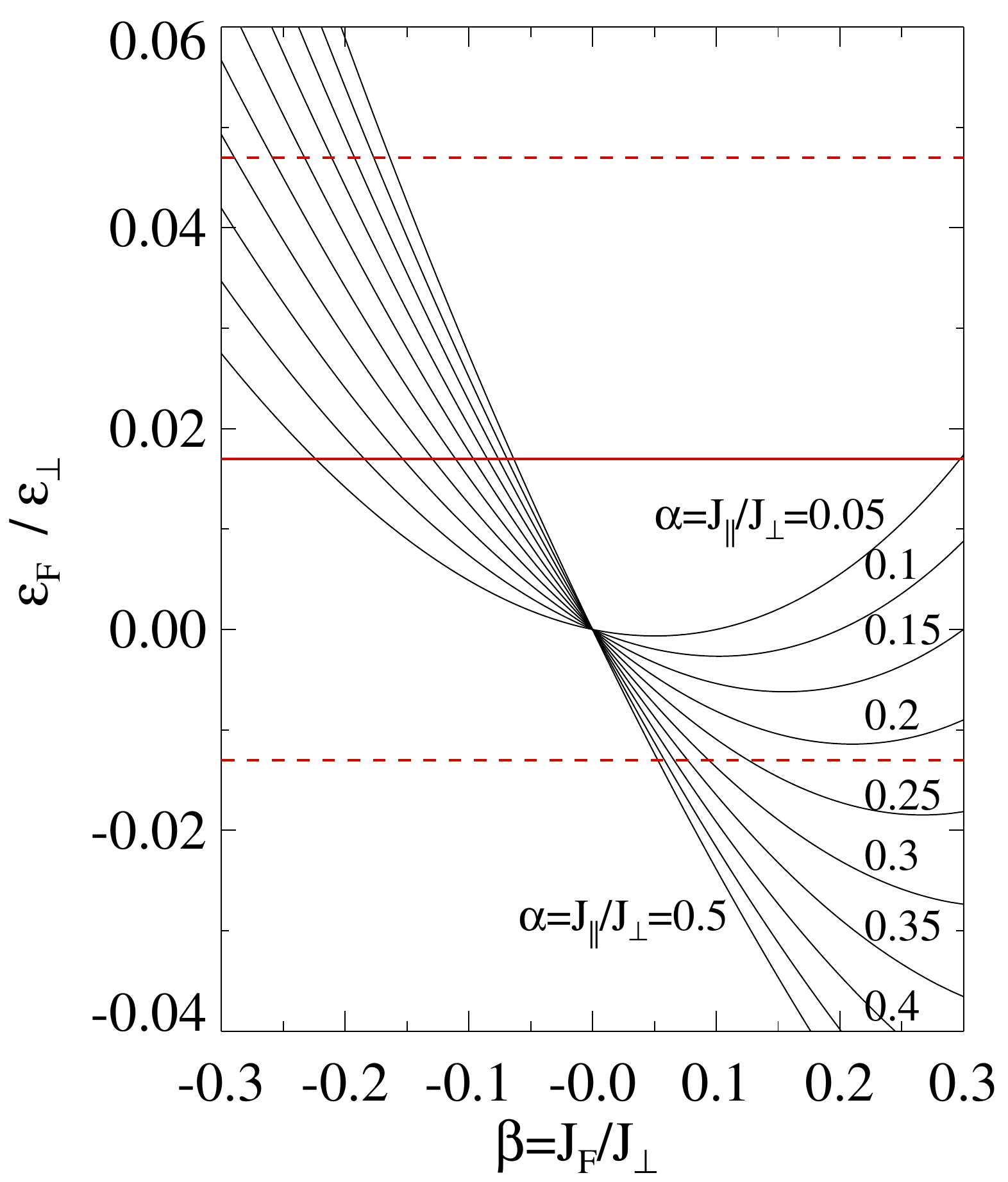}%
\caption{\label{SS} (color online) Solid black lines show bond energy ratios for $J_F$ and $J_\bot$,
as a function of $\beta=J_F/J_\bot$, for several values of $\alpha=J_\|/J_\bot$, from 0.05 to 0.5, every 0.05, as estimated by the
continuous unitary transformation. The experimental value and corresponding errors are
shown as horizontal lines. }
\end{center}
\end{figure}
\begin{figure}[!h]
\begin{center}
\includegraphics[width=3.3in,angle=0]{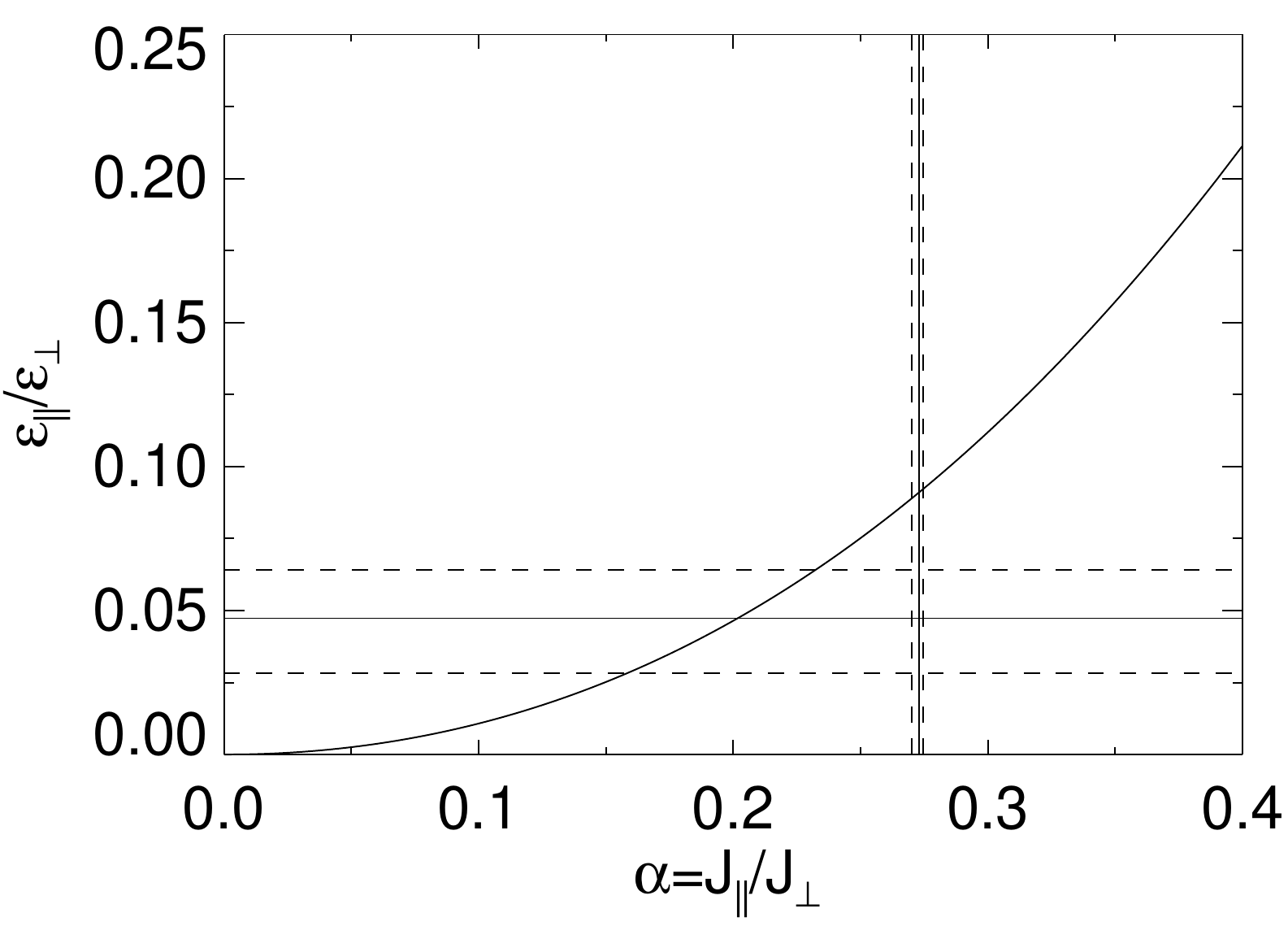}%
\caption{\label{bondEladder} Ratio of leg to rung bond energies versus $\alpha=J_\|/J_\bot$,
as calculated using the continuous unitary transformation. Experimental values are shown as
horizontal/vertical solid lines, with error-bars indicated by dashed lines.}
\end{center}
\end{figure}

Theoretical bond energies were obtained using a particle conserving continuous unitary transformation
(CUT).\cite{Knetter00, Knetter03EPJ, Knetter03Jphys} The CUT is realized perturbatively at the isolated
rung dimer limit. The elementary excitations conserved after the transformation are triplons.\cite{Schmidt03}
The static correlation functions for perpendicular, parallel, and diagonal bonds of the two-leg ladder
can then be determined from the ground-state energy per bond, $E_0 (J_{\perp},J_{\parallel},J_{F})/N_{\rm bonds}$,
which we calculated exactly up to order 7 in $J_{\parallel}/J_{\perp}$ and $J_{F}/J_{\perp}$.
Note that we are using the bare series in the following since we restrict the discussion to small and
intermediate values of the couplings $J_{\parallel}$ and $J_{\rm F}$.
Using the Feynman-Hellman theorem, one finds for the static dimer correlation function
\begin{equation}
C_{\perp} = \langle {\bf S}_{j,1} {\bf S}_{j,2} \rangle = \frac{1}{N_{{\rm bonds},
\perp}}\frac{\partial}{\partial J_{\perp}} \langle \hat{\cal H} \rangle \;\;\;,
\end{equation}
with analogous expressions for other bonds.

For comparison to the experimental data, calculations of
$\varepsilon_{F} / \varepsilon_{\perp}$ versus $\beta=J_F/J_\perp$ are presented for various values of
$\alpha=J_\|/J_\perp$ in Fig.~\ref{SS} (black solid lines). The experimental result
$\varepsilon_{F}/\varepsilon_\bot $ is shown as a red line in Fig.~\ref{SS},
with the dashed lines indicating uncertainty. This measurement does not impose additional constraints on $J_F$.

For an ideal ladder, without frustrating or inter-ladder interactions, the leg to rung bond energy ratio  versus $\alpha$, computed by CUTs,
is shown in Fig.~\ref{bondEladder}. The bond energy ratio extracted from the neutron scattering
data is shown with dashed lines. Given the rough nature of our approximation to the covalent form factor, the level of agreement is acceptable.

\subsubsection{Inter-ladder exchange}

\begin{figure}[!h]
\begin{center}
\includegraphics[width=3.3in,angle=0]{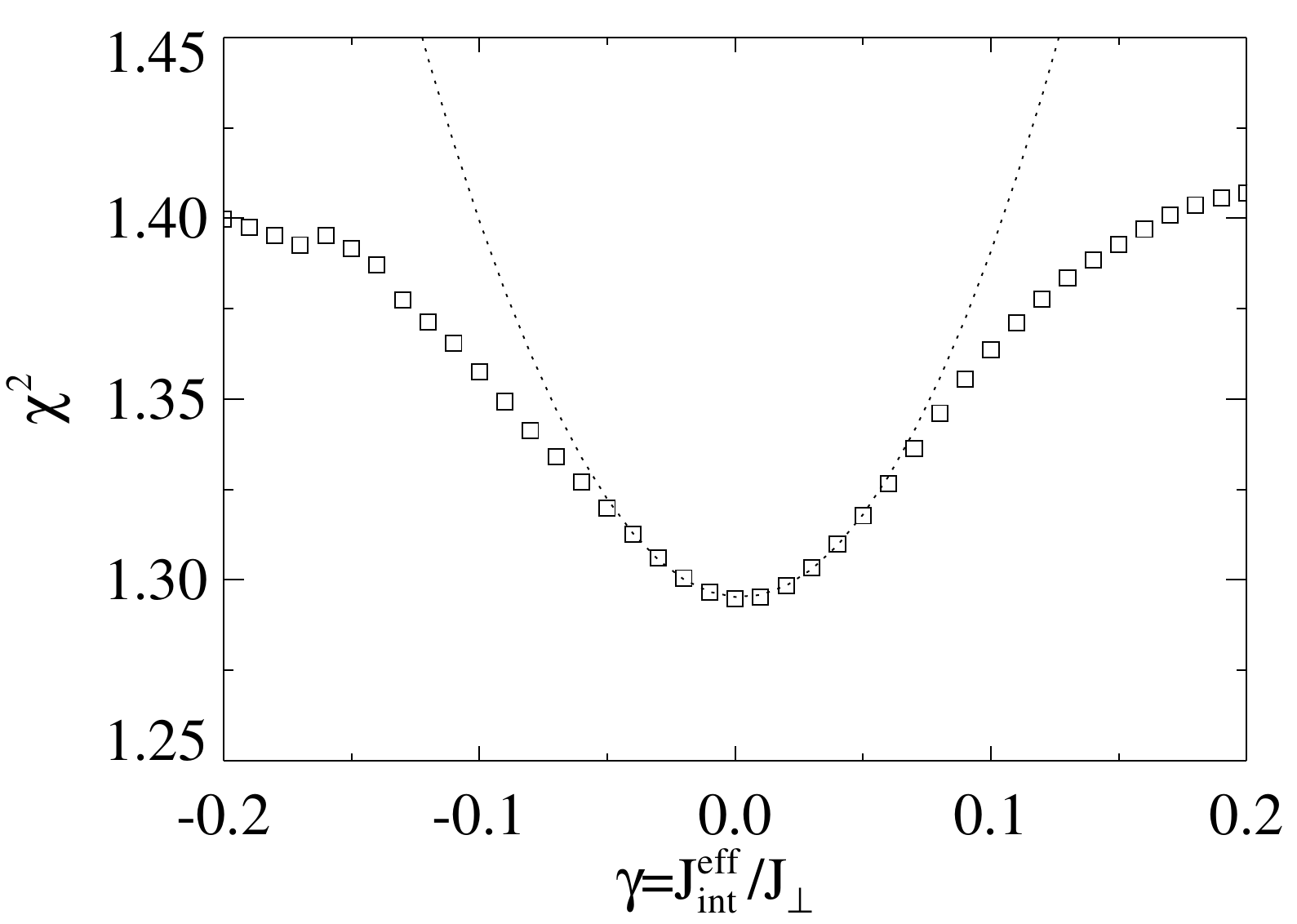}%
\caption{\label{chiofl} Reduced $\chi^2$ for the global fit to BCPB inelastic neutron scattering
data as a function of $\gamma=J_{\rm int}^{\rm eff}/J_\bot$. The dashed line is a quadratic fit for points around the minimum,
which indicates that $\gamma = 0.002 \pm 0.006$.}
\end{center}
\end{figure}

The fit used in the previous section finds that $J_{\rm int}^{\rm eff}$ is at least two orders of
magnitude smaller than $J_\|$ and $J_\bot$.
To evaluate the robustness of this finding and obtain a quantitative uncertainty limit on $\gamma$,
the fit was repeated for several different fixed values of $\gamma$.  The resulting values for the
reduced $\chi^2$ are plotted versus $\gamma$ in Fig.~\ref{chiofl}. A quadratic fit close to $\gamma=0$,
yields $\gamma = 0.002 \pm 0.006$, and this analysis indicates at least two orders of magnitude difference
between $J_{\rm int}^{\rm eff}$ and $J_\|$. Note that this result is specific to the assumed nature of inter-ladder dispersion.\cite{Matsumoto04}

\begin{figure}[]
\begin{center}
\includegraphics[width=3.3in,angle=0]{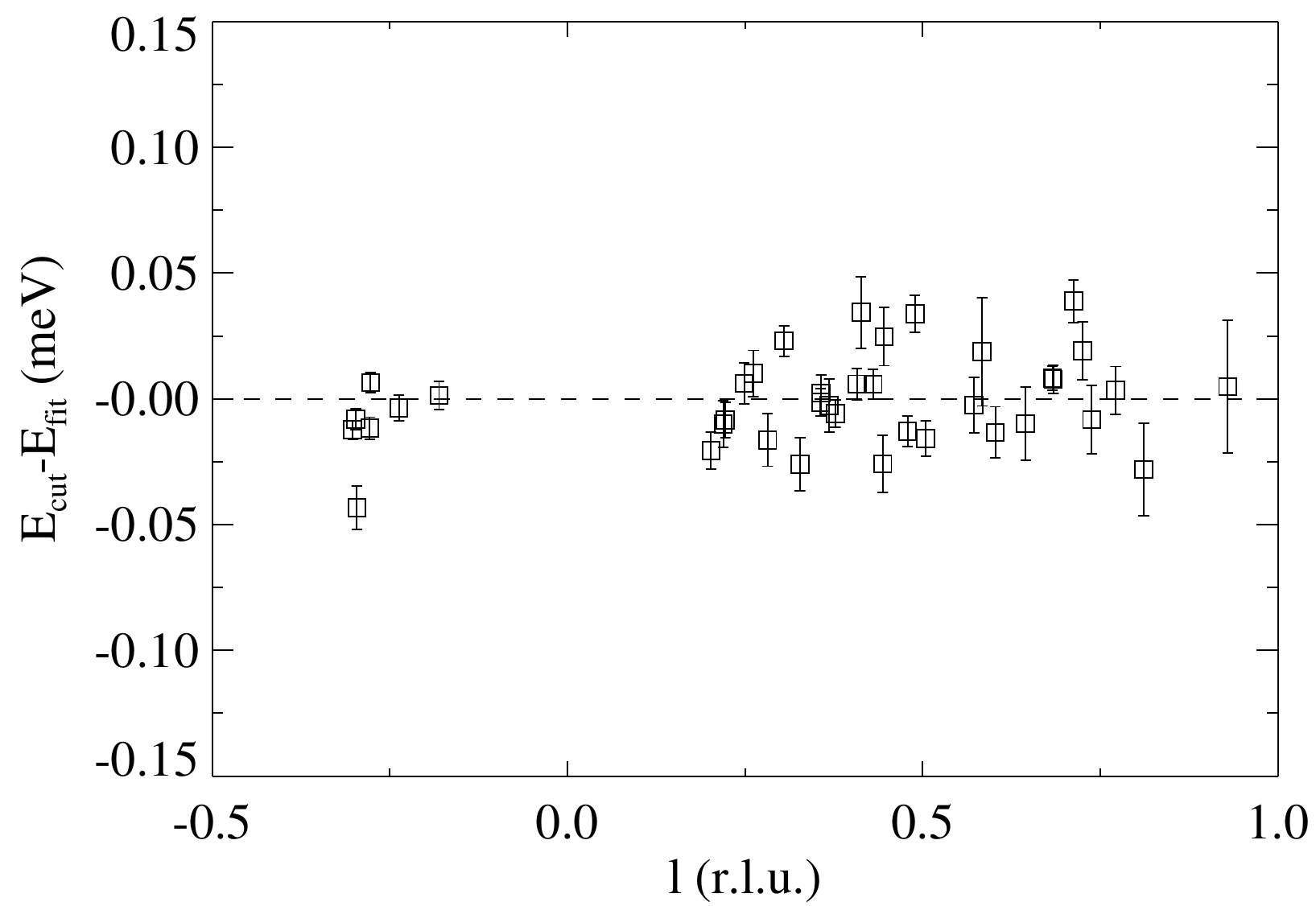}%
\caption{\label{rez} The difference between the best fit one dimensional dispersion relation and gaussian fits to cuts through raw data resolved versus wave vector transfer along ${\bf c}^*$.   No systematic $l$ dependence is observed indicating that to within error a one dimensional model is appropriate for BPCB ($\gamma = 0.002 \pm 0.006$).}
\end{center}
\end{figure}

A second, less model dependent approach, involves generating constant
{\bf q}-cuts (width 0.05 r.l.u.) through the experimental data at different $h$ values for all three sample
orientations.  The difference between the Gaussian fitted peak position to such cuts and the strictly
one-dimensional dispersion relation is plotted as a function of $l$ in Fig.~\ref{rez}.
The error-bars in the figure are the positional uncertainties of these Gaussian fits and should be
compared to the energy resolution of the instrument, $\Delta E \sim 0.1$ meV. For some cuts, no signal
was observed above background, in which case no point is shown in Fig.~\ref{rez}.  There is no apparent
systematic deviation from a zero residual as a function of $l$, again indicating the absence of
magnon dispersion perpendicular to ${\bf a}^*$.


Low temperature NMR measurements by Klanj\v{s}ek {\it et al.}\cite{Klanjsek08} reveal
3D magnetic order for $T <100$~mK. From this observation, an average inter-ladder coupling of
$\sim~20$~mK ($\sim~1.7~\mu$eV) was inferred, and an identical result was found using
neutron diffraction at high magnetic fields.\cite{Thielemann08}
These results considered four
nearest neighbors in their mean field expansion.  Therefore, the strength of the total effective inter-ladder
exchange energy is $\sim~80$~mK. This value is comparable to the limits on
$|J_{\rm int}^{\rm eff}| \lesssim~70$~mK set by our fit.

\section{Conclusions}
We have shown that (C$_5$D$_{12}$N)$_2$CuBr$_4$ is an excellent realization of two-leg spin-1/2 ladder
in the strong coupling limit.  The inferred rung exchange $J_\bot=(1.09\pm0.01)$~meV and
leg exchange $J_\|=(0.296\pm0.005)$~meV are in excellent agreement with values obtained from other techniques.\cite{Watson01,Klanjsek08,Lorenz08,Anfuso08}
Using two different methods of analysis, we showed that the effective inter-ladder exchange $J_{\rm int}^{\rm eff}$ is more than
two orders of magnitude smaller than $J_\bot$.
These results confirm the previous conclusion that BPCB can be
classified as a one-dimensional system.
Alone, the neutron data do not provide a direct
measurement of $J_F$. However, in combination with high field magnetization studies, NMR measurements, and
theoretical calculations of the dispersion relation, the neutron data sets an upper limit on
$J_F+J_{F'}$, which is one order of magnitude smaller than the rung exchange ($|(J_F+J_{F'})/J_\bot| \leq 0.1$).

The single mode approximation provides an excellent account of the data and the intra-ladder bond energies extracted
are in agreement with results from continuous unitary transformations within experimental error. The intensity pattern can be understood only if covalency effects are taken into account.
The experiment elucidates the spin interactions in BPCB for analysis of recent\cite{Thielemann09}
and future high field neutron scattering experiments.

During preparation of this manuscript, we became aware of the field dependent measurements of Thielemann {\it et al.}\cite{Thielemann09}  Our data at zero field confirm their results, adding information regarding frustrating interactions and bond energies.

\begin{acknowledgments}
The authors thank T.~E.~Sherline and J.~R.~D.~Copley for technical assistance during the experiment and
D.~A.~Jensen for generating the samples. We also thank Y.~Qiu for help in extracting data from the
DAVE software package. We acknowledge helpful comments from C. Ruegg. The NSF funds the NCNR under agreement No. DMR-0454672, CLB and ATS through
DMR-0603126, DMP and MWM through DMR-0701400, DRT through DMR-0453362, and the NHMFL via cooperative agreement
DMR-0654118. ORNL is managed by UT-Battelle, LLC,
for the U.S.~Department of Energy under contract DE-AC05-00OR22725. KPS acknowledges ESF and
EuroHorcs for funding through his EURYI. GSU acknowledges the support of the
Heinrich-Hertz Stiftung NRW for his leave. ATS appreciates the hospitality afforded to him as a
visiting scientist in the Neutron Scattering Sciences Division at ORNL.
\end{acknowledgments}
\newpage
\bibliography{Savici-BPCB}

\end{document}